\documentclass[11pt,a4paper]{article} 
\pdfoutput=1
\usepackage{jcappub} 

\title{Dark Matter in a Bouncing Universe} 

\author[1]{Yeuk-Kwan~E.~Cheung,}

\author[1, 2]{Jin~U~Kang }

\author[1]{and Changhong~Li }  

\affiliation[1]{Department of Physics, Nanjing University,\\ 
22 Hankou Road, Nanjing, China 210093} 

\affiliation[2]{Department of Physics, Kim Il Sung University, \\RyongNam Dong, TaeSong District, Pyongyang, DPR. Korea}

\emailAdd{cheung@nju.edu.cn} 
\emailAdd{jin.u.kang2@gmail.com} 
\emailAdd{chellifegood@gmail.com}

\abstract{We investigate a new scenario of dark matter production in a bouncing universe, in which dark matter was produced completely out of equilibrium in the contracting as well as expanding phase. We explore possibilities of using dark matter as a probe of the bouncing universe, focusing on the relationship between a critical temperature of the bouncing universe and the present relic abundance of dark matter.
} 
\keywords{Dark Matter, Bouncing Universe} 
\arxivnumber{0000.0000} 
\begin{document} 
\maketitle

\section{Introduction}
\label{sec:intro}

Bouncing cosmology aims at resolving the cosmological singularity, which is one of the most important problems of the cosmology (see~\cite{Novello:2008ra, Brandenberger:2012zb} for recent reviews), mostly relying on more fundamental theories of quantum gravity, e.g. loop quantum gravity or string theory (see e.g. \cite{Loewenfeld:2009aw}). Furthermore, as alternative to inflationary models, there are many proposed bouncing models that attempt to solve the other cosmological problems that inflationary cosmology does, but without invoking early accelerated expansion. In past years, detailed realizations of working bounce, which aim to pridict a stable and scale-invariant power spectrum compatible with the current observations~\cite{Komatsu:2010fb,Planck:2013kta}, have been hotly contested 
(see e.g. ~\cite{Wands:1998yp, Khoury:2001wf, Gasperini:2002bn, Creminelli:2006xe, Cai:2007qw, Cai:2008qw,  Wands:2008tv,  Li:2011nj, Lilley:2011ag, Easson:2011zy,  Bhattacharya:2013ut,  Li:2013bha}).
In light of the observations of BICEP2, various predictions from the bouncing universe have been discussed \cite{Li:2014msi, Quintin:2014oea, Liu:2014tda, Li:2014cka, Wan:2014fra,Li:2014qwa,Cai:2014hja,Cai:2014xxa,Hu:2014aua,Cai:2014zga, Lindley:2014gia}, and most notably, the tensor-scalar ratio of a particular model in bouncing cosmology has been worked out to fit to the WMAP and BICEP2 data~\cite{Xia:2014tda}.  However, the bouncing cosmology is still a controversial subject \cite{Battefeld:2014uga, Kallosh:2007ad}. 


Dark matter is one of the most prominent objects in cosmology and particle physics (for recent reviews and many  references, see \cite{Bertone:2010zza}).  Dark matter constitutes substantial proportion of the energy density of the current Universe and is supposed to play a crucial role in cosmology. The standard lore is that dark matter is the thermal relic which was initially in full thermal (kinetic and chemical) equilibrium and then froze out via thermal decoupling \cite{Scherrer:1985zt, Gondolo:1990dk, Feng:2008ya}. Since dark matter was assumed to be created in the very early times there were interesting attempts to use dark matter as a probe to constrain the very early universe or non-standard cosmology \cite{Kang:2008zi, Kang:2008jq, Drees:2007kk}.           

In this short note we discuss a scenario that the thermal production of dark matter can be extended  beyond the big bang in bouncing universe, which was first put forward in \cite{Li:2014era}.  The aim of this work is not to propose new bouncing scenarios, but to explore possibilities of proving bouncing universe in terms of dark matter.    
To this end we rely on mechanism in which the dark matter is out of chemical equilibrium when it was being created. The reason is that in the standard thermal decoupling approach the relic density is determined by the abundance at the freeze-out temperature, so it is impossible to extract any information about the earlier phase before the freeze-out time. Therefore we invoke this out-of equilibrium scenario to explore the interplay between dark matter and the bouncing universe. 
We find that the relic abundance of dark matter depends on some temperature,  above which new physics regime responsible for the bounce takes over the universe, and this relationship can be used to probe the bouncing universe. An important assumption is made that the duration of bouncing phase, in which non-standard physics is dominant, is sufficiently short not to change the number of dark matter particles per comoving volume. This assumption makes possible a model-independent study, in which details of realization of bounce are not relevant, so we do not address any explicit bouncing dynamics in this paper.        
To summarize, we show that the scenario, in which  dark matter particles are produced out of equilibrium in pre-bounce contraction as well as the post-bounce expansion epoch,  opens up a new possibility to satisfy the currently  observed relic abundance and this study can be used to draw some information about bouncing universe. 

The remainder of this paper consists of three sections. In section \ref{sec:bas},  we set up the basic scenario for dark matter production in bouncing universe and present formalisms to compute the relic density of dark matter. In section \ref{sec:ill}, we discuss two simple cases to illustrate how the relic density of dark matter depends on the characteristic temperature of bounce. We conclude in the last section.   

\section{
Out-of-equilibrium production of dark matter in bouncing universe}
\label{sec:bas}


The time evolution of number density $n_\chi $ of dark matter particle $\chi$ is governed by the following Boltzmann equation \cite{Kolb:1990vq}:
\begin{equation} \label{nsf}
\frac{dn_\chi}{dt}+ 3 H n_\chi= -\langle \sigma v \rangle (n_\chi^2-n_{\chi,{\rm EQ}}^2), 
\end{equation}
where 
$H$  and $n_{\chi,{\rm EQ}}$ are the Hubble parameter and the equilibrium number density of $\chi$, respectively.  $\langle \sigma v \rangle $ is the thermal average of the total annihilation cross section times relative velocity.  
By introducing dimensionless quantities\footnote{In this paper we use natural units, where $\hbar=c=k_B=1$.} 
$x=m_\chi/T$, $Y= n_\chi/s$ and $Y_{\rm EQ} = n_{\chi,{\rm EQ}}/s$ with temperature $T$ and entropy density $s$, Eq. \eqref{nsf} can be written as
\begin{equation} \label{Y}
\frac{dY}{dx}= -\frac{s}{x H}\langle \sigma v\rangle(Y^2-Y_{\rm EQ}^2). 
\end{equation} 
The entropy density is given by $s=(2 \pi^2/45)h_* T^3$ with $h_*$ being the relativistic degrees of freedom for entropy density, while the Hubble parameter during the radiation dominated era is $H=\frac{\pi T^2}{M_p}\sqrt{\frac{g_*}{90}} $ with $M_p$ being the reduced Planck mass and $g_*$ the relativistic degrees of freedom for energy density.  Throughout we take  $h_* \simeq g_*$ during the phases under consideration. 

In the standard thermal decoupling approach, the dark matter particles are assumed to have been in full equilibrium and its abundance tracks equilibrium value until thermal decoupling, which occurs around freeze-out temperature.  In this case the relic density is essentially determined by the equilibrium abundance at freeze-out temperature and this is why any information of the earlier phase is washed out. Furthermore the relic density is inversely proportional to the cross section, i.e.
$\Omega_\chi  \propto 1/\langle\sigma v\rangle$. 

On the other hand,  dark matter production in the non-chemical equilibrium was also studied in the standard expanding universe \cite{Chung:1998ua, Drees:2006vh}. In this approach it is assumed that the cross section is so small that the dark matter particles were completely out of chemical equilibrium when they were being produced. The feature of this non-equilibrium approach is that the relic density is proportional to the cross section, i.e. $\Omega_\chi  \propto \langle\sigma v\rangle$, in contrast to the standard equilibrium approach and that there is no freeze-out temperature. Thus this approach is able to probe the very early universe and has been used to constrain the non-standard cosmolgy~\cite{Drees:2007kk}. 
For this reason we adopt this non-chemical equilibrium mechanism of dark matter production (but kinetic equilibrium is preserved).  In order for dark matter not to attain the chemical equilibrium, the cross section $\langle\sigma v\rangle$ should be very small. Below we assume that the actual abundance of dark matter has been always much less than the equilibrium value, so we will ignore the $Y^2$ term on the right hand side of Eq. \eqref{Y}. 

Now we combine this out-of-equilibrium mechanism of dark matter production with cosmic evolution of bouncing universe. 
We consider the bouncing universe with following three phases: 
\begin{itemize}
\item[]{{\bf Phase I}:
the pre-bounce contraction phase ($H<0$). 
During this phase the universe  is dominated by thermalized radiational background getting hotter with contraction. The contracting radiation-dominated phase ends at temperature $T_b^-$, at which non-standard physics starts to become dominant to give rise to bounce. 
}\item[]{{\bf Phase II}:
the post-bounce expansion phase ($H>0$). After bounce the universe recovers standard radiation dominated era with maximal temperature $T_b^+$, at which phase II begins. This phase corresponds to standard hot big bang phase. $T_b^+$ is analogous to the reheating temperature in the inflationary cosmology. For simplicity we assume  $T_b^+\simeq T_b^-$ and set $T_b^{\pm}=T_b$ in the following.  $T_b$ can be viewed as a critical temperature of the phase transition from standard physics regime into new physics one, which is responsible for the bounce. 
} \item[]{{\bf Phase III}:
bouncing phase, in which non-standard dynamics such as e.g. ghost condensate gives rise to bounce. This phase connects phase I to II. We assume no production of entropy during bounce, so entropy is conserved through this phase. 
}  \end{itemize}

We assume that the bounce occurred very quickly. Namely the characteristic time of bounce (i.e. duration of phase III) is much shorter than dark matter reaction time scale (inverse of the dark matter annihilation rate $\Gamma_\chi=n_\chi \langle\sigma v\rangle$), so that the phase III is too short to produce/annihilate dark matter\footnote{The assumption of short duration of bounce was also made in  bouncing scenarios, which require a violation of the null energy condition, in order for instabilities not to develop, see e.g. \cite{Buchbinder:2007ad, Kallosh:2007ad}.}. In other words the number of dark matter particles per comoving volume changes little cross bouncing phase. Together with entropy conservation this implies that the abundance of dark matter remains almost constant over this phase and one can take the abundance at the end of phase I to be the initial abundance of phase II. Due to this matching condition we can leave out phase III from consideration and will focus on phase I and II in what follows. Throughout this paper we make use of superscript $-$ and $+$ to refer to contracting (phase I) and expanding phase (phase II) respectively, wherever the distinction is necessary. 
For instance $Y^{-}$ and $Y^{+}$ stand for the abundances in contracting (phase I) and expanding phase (phase II), respectively.
The matching condition mentioned above reads  $Y^{-} (x_b^{-}) \simeq Y^{+} (x_b^{+})$, where $x_b^{\pm}=m_\chi/T_b^{\pm} \simeq m_\chi/T_b = x_b$.   

To sum up one should integrate ($f= 1.32 \sqrt{g_*}m_\chi M_{p} $)
\begin{equation} \label{Y1}
\frac{dY^{\pm}}{dx} = \mp f \frac{\langle \sigma v\rangle}{x^2}\left(\left(Y^\pm\right)^2- Y_{\rm EQ}^2\right) 
\end{equation}
under assumption $Y^{\pm}\ll Y^{\rm EQ}$ to obtain the relic abundance $Y_\infty \equiv Y^+(x \gg x_b)$ using 
conditions\footnote{We assume that the dark matter abundance in the far past of the contracting phase was negligible.} 
\begin{equation} \label{cond}
Y^-(x \gg x_b)\simeq0, \quad Y^-(x_b)\simeq Y^+(x_b).
\end{equation} 
The overall sign $\mp$ on the right hand side of Eq. \eqref{Y1} corresponds to the sign of Hubble parameter. Then the present relic density is 
\begin{equation} \label{Omega}
\Omega_\chi h^2 = 2.8 \times 10^8\, m_\chi Y_\infty {\rm GeV}^{-1},
\end{equation}
which is approximately 0.1 to account for the observational bound on the cold dark matter \cite{Lahav:2014vza}. $h$ is the scaled Hubble constant, $h\simeq0.7$. 

\section{Two illustrative cases: low- and high-temperature cases}
\label{sec:ill}
Now we concentrate on two extreme cases, i.e. low-temperature bounce ($T_b \ll m_\chi$) and high-temperature bounce ($T_b \ll m_\chi$), for which tractable approximate calculations of analytic results are allowed. 
     
\paragraph{\bf Low-temperature bounce: $T_b \ll m_\chi$ (i.e. $x_b \gg 1$).}

In this case  $\chi$ particles are non-relativistic in phase I and II ($x\gg1$), for which we have $Y_{\rm EQ}^2=c\, x^3 e^{- 2x}$ with $c=0.021 g_\chi^2/g_*^2$ and $g_\chi$ being the number of degrees of freedom of $\chi$. In the non-relativistic case the thermal average of the annihilation cross section can be expanded as $\langle\sigma v\rangle= a + \frac{6b}{x} + O\left(\frac{1}{x^2}\right)$ (see 
\footnote{Note that there are model-dependent exceptions where this approximation for the thermal average of annihilation cross section is not valid.  For instance, this form can not be used 
in case of so-called {\it coannihilations} that occur if some other particles have a mass similar to the relic particle and share a quantum number with it (see e.g. \cite{Griest:1990kh}). 
We restrict our consideration to cases where those exceptions do not arise, as a main purpose of this paper 
is to illustrate an idea of using dark matter to probe the energy scale of a bounce in a model independent form.}
e.g. \cite{Gondolo:1990dk}) and we approximate  $\langle\sigma v\rangle\simeq a$ for simplicity. 
Then by integrating Eq. \eqref{Y1} from $x_i^\pm$ to $x$ with $\left(Y^\pm\right)^2$ dropped out on the right hand side gives the solution
\begin{equation}
Y^{\pm}(x) \simeq \pm \frac{1}{2}\,f\,c\,a (x_i^\pm e^{-2 x_i^\pm} - x e^{-2 x }) + Y_i^{\pm},     
\end{equation}
where subscript $i$ stands for initial value,  e.g. $Y_i^{\pm}$ is the initial value of $Y^{\pm}$, i.e. $Y_i^{\pm}=Y^{\pm}(x_i^{\pm})$.  Note that $x_i^-\gg x$ in phase I and $x_i^+=x_b$, so the condition \eqref{cond} entails $Y_i^-\simeq0$ and $Y_i^+ \simeq Y^-(x_b)$.  
Thus we obtain
\begin{eqnarray}
&&Y^-(x_b) \simeq 0.014g_\chi^2g_*^{-3/2}m_\chi M_p\, a\,x_b\,e^{-2 x_b}, \label{L} \\
&&Y_\infty=Y^+(x\gg x_b) \simeq 2 Y^-(x_b). 
\end{eqnarray} 
Comparing to the result of low-temperature scenario in the standard expanding universe (see (13) in \cite{Drees:2006vh}), 
where our $x_b$ corresponds to $x_0$ (note that in that scenario it was assumed that the dark matter started to be thermally produced at $x_0$, i.e. $Y(x_0)=0$), the abundance is enhanced by factor 2. This enhancement is due to the contribution from contracting phase before bounce in addition to expanding phase.   


\paragraph{\bf High-temperature bounce: $T_b \gg m_\chi$ (i.e. $x_b \ll 1$).}

In this case we can assume that $\chi$ particles are mostly created while being relativistic.  For the relativistic particles  $Y_{\rm EQ}$ is independent of $x$ (or temperature), i.e. $Y_{\rm EQ}\simeq0.278\,g_{\rm eff}/g_*$ with $g_{\rm eff} = g_\chi$ for boson and $g_{\rm eff} = 3g_\chi/4$ for fermion. We use $\langle\sigma v \rangle \simeq \sigma_0\, x^{-n}$ with $n\geq0$ (typically $n=2$ for the pair annihilation processes of Dirac and Majorana fermions into a pair of massless fermions). Then it is easy to integrate the Boltzmann equation \eqref{Y1} (neglecting $\left(Y^{\pm}\right)^2$ again on the right hand side) to get solution
\begin{equation}
Y^{\pm}(x) \simeq \pm \frac{ 0.077\,g_{\rm eff}^2\,f\,\sigma_0}{g_*^2 (n+1)} \left(\frac{1}{(x_i^\pm)^{n+1}}  
- \frac{1}{x^{n+1}} \right) + Y_i^{\pm}.     
\end{equation} 
Applying the boundary condition \eqref{cond} again we obtain
\begin{eqnarray}
&&Y^-(x_b) \simeq  \frac{0.102\, g_{\rm eff}^2 g_*^{-3/2}m_\chi M_{p}\,\sigma_0}{(n+1)\,x_b^{n+1}}, \label{H} \\
&&Y_\infty=Y^+(x\gg x_b) \simeq 2 Y^-(x_b).
\end{eqnarray}

\paragraph{\bf Condition for non-chemical equilibrium.}

Let us consider to what extent our anstaz for non-chemical equilibrium could be valid. To this end let us rewrite the Boltzmann equation \eqref{Y} as
\begin{equation}
\frac{x}{Y}\frac{dY}{dx}= \frac{n_{\chi} \langle \sigma v\rangle}{H}\left(\frac{Y_{\rm EQ}^2}{Y^2}-1\right)
\end{equation}  
From this equation it is clear that the sufficient condition for non-equilibrium (i.e. $Y_{\rm EQ}^2/Y^2>1$) is $n_{\chi} \langle\sigma v\rangle < |H|$. Therefore if the dark matter was produced mainly at some temperature $T_*$ with $\Omega_\chi<1$, the condition becomes
$10^{10} ({\rm GeV})^2\langle \sigma v\rangle <x_*$  with $x_*=m_\chi/T_*$(cf. (1) in \cite{Chung:1998ua}). Since in our case the dominant contribution comes from $T_b$ (or $x_b$), we obtain the condition for non-chemical equilibrium 
\begin{equation} \label{NEQ}
10^{10} ({\rm GeV})^2\langle \sigma v\rangle <x_b.
\end{equation} 
Since $x_b\gg1$ for low-temperature case and $x_b\ll1$ for high-temperature one, this condition requires much smaller cross section for the latter than for the former.  The above condition constrains temperature $T_b$ as well as cross section. For instance in the case of high-temperature bounce, for given cross section and $m_\chi$ the temperature $T_b$ can not be arbitrarily high in order to satisfy this non-chemical equilibrium condition. 

\paragraph{\bf Present relic density and observational constraint.}

For both cases considered above the present relic density \eqref{Omega}  becomes
\begin{equation} \label{Omega1}
\Omega_\chi h^2 \simeq 5.6 \times 10^8\, m_\chi Y^-(x_b)\, {\rm GeV}^{-1},
\end{equation}
where $Y^-(x_b)$ is given by \eqref{L} and \eqref{H} in two cases, respectively. 

\begin{figure}[htp] 
\centering
\includegraphics[width=0.8\textwidth]{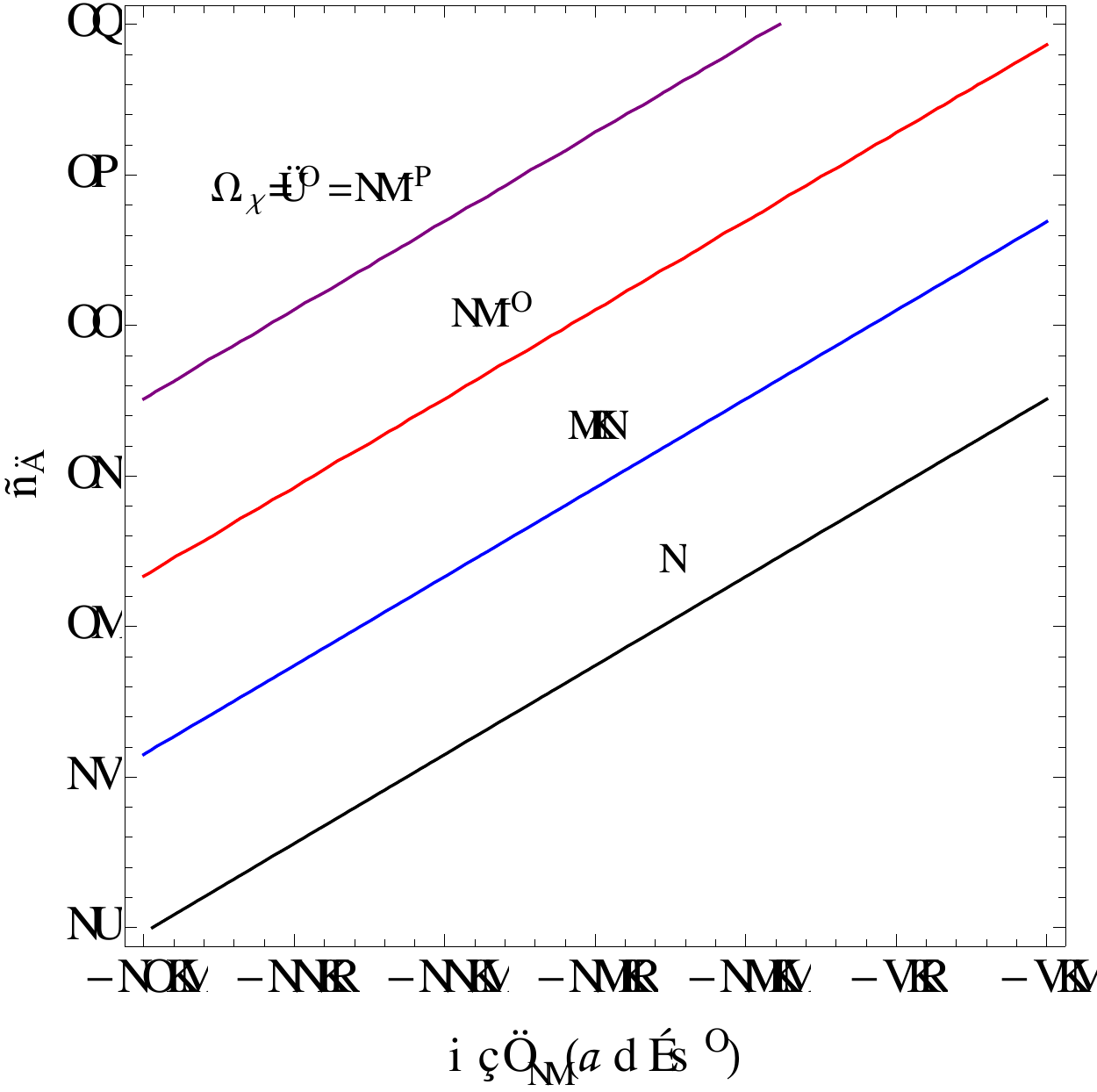}
\caption{Low-temperature case: contour plots of the predicted relic abundances in the ($x_b$, $a$) plane. Here we take $\langle\sigma v \rangle \simeq a$ and $m_\chi=100{\rm GeV}$. }\label{fig:fig_low_a-x-Omega}
\end{figure}

\begin{figure}[htp] 
\centering
\includegraphics[width=0.8\textwidth]{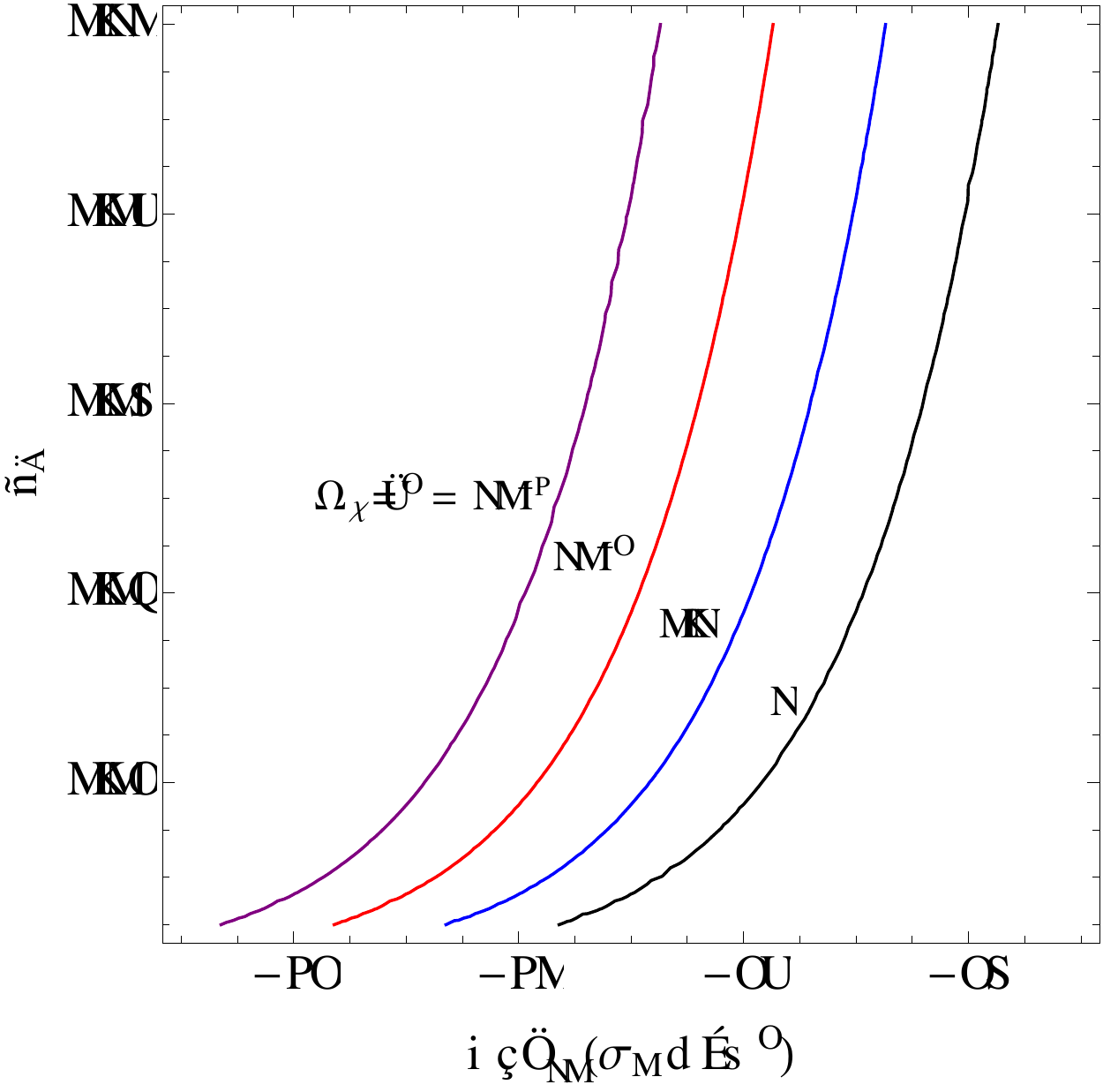}
\caption{High-temperature case: contour plots of the predicted relic abundances in the ($x_b$, $\sigma_0$) plane. Here we take $m_\chi=1{\rm GeV}$ and $n=2$ for $\langle\sigma v \rangle \simeq \sigma_0\, x^{-n}$.}\label{fig:fig_high_a-x-Omega}
\end{figure}

\begin{figure}[htp] 
\centering
\includegraphics[width=0.8\textwidth]{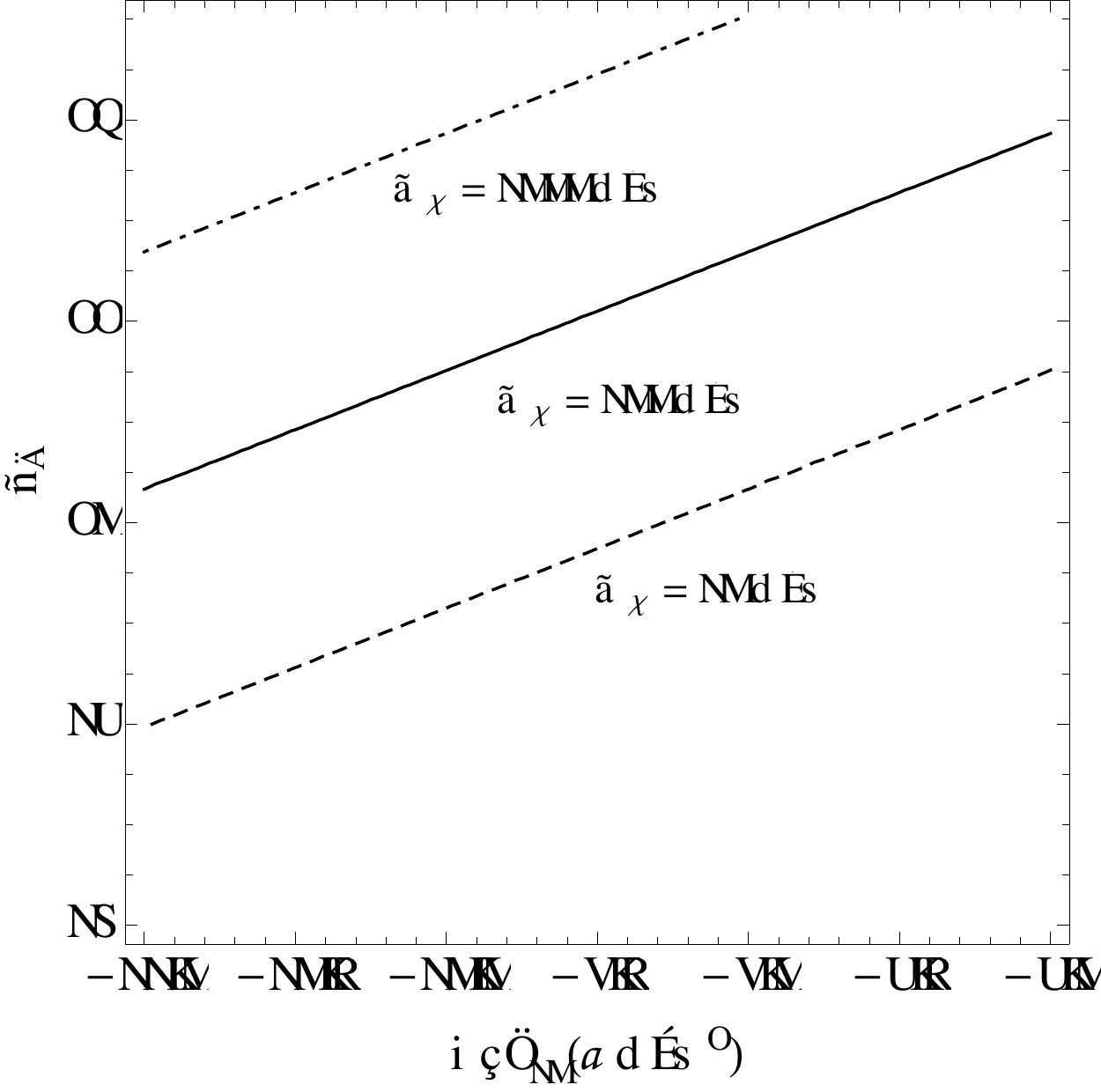}
\caption{Low-temperature case: $x_b$ as a function of the cross section  $\langle\sigma v \rangle \simeq a$ to reproduce the present relic abundance $\Omega_\chi h^2\simeq0.1$ for various $m_\chi$.}\label{fig:fig_low_a-x-m}
\end{figure}

\begin{figure}[htp] 
\centering
\includegraphics[width=0.8\textwidth]{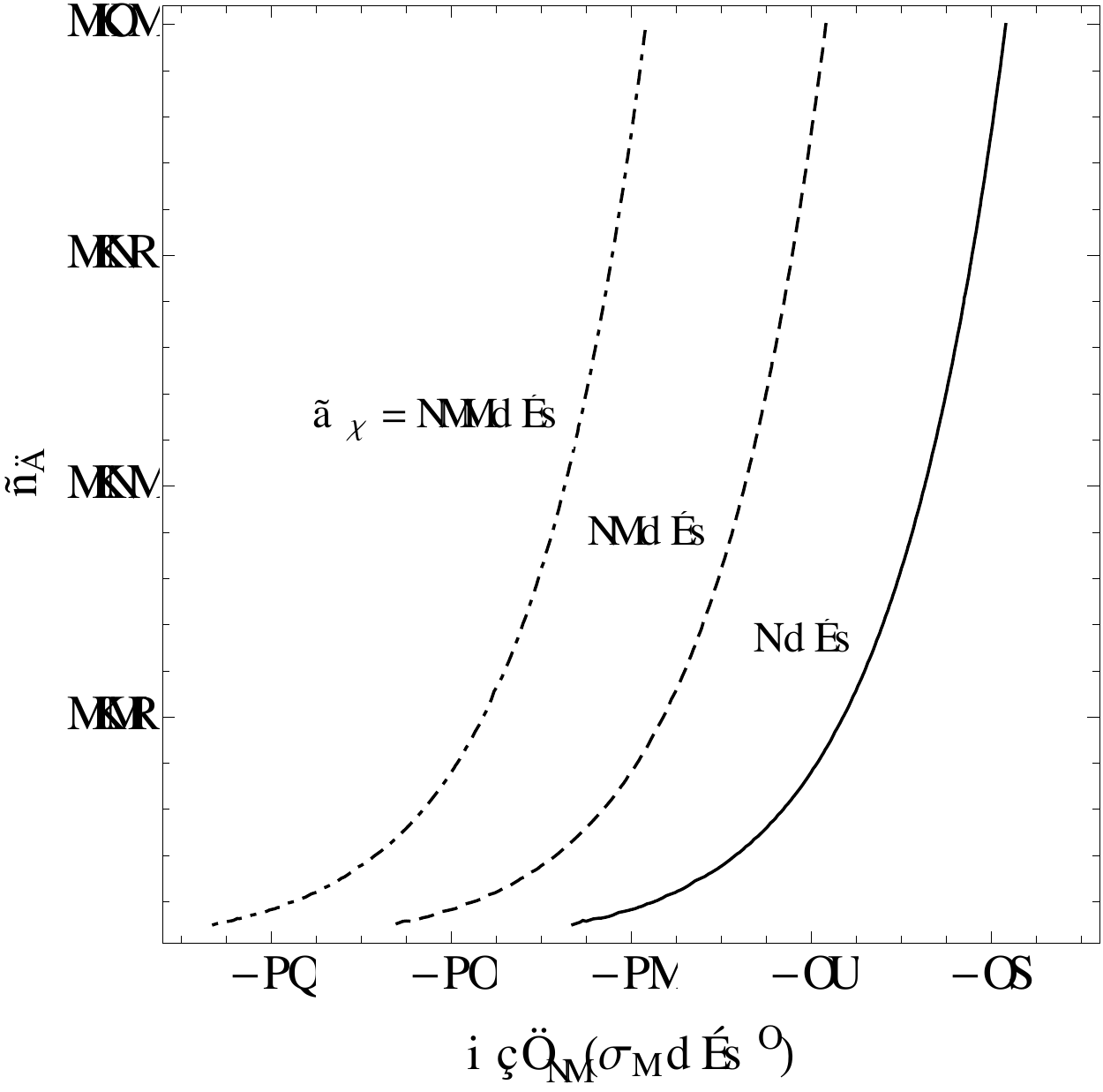}
\caption{High-temperature case: $x_b$ as a function of  $\sigma_0$ to reproduce the present relic abundance $\Omega_\chi h^2\simeq0.1$ for various $m_\chi$. Here we take $n=2$ for the cross section $\langle\sigma v \rangle \simeq \sigma_0\, x^{-n}$}\label{fig:fig_high_a-x-m}
\end{figure}

\begin{figure}[htp] 
\centering
\includegraphics[width=0.8\textwidth]{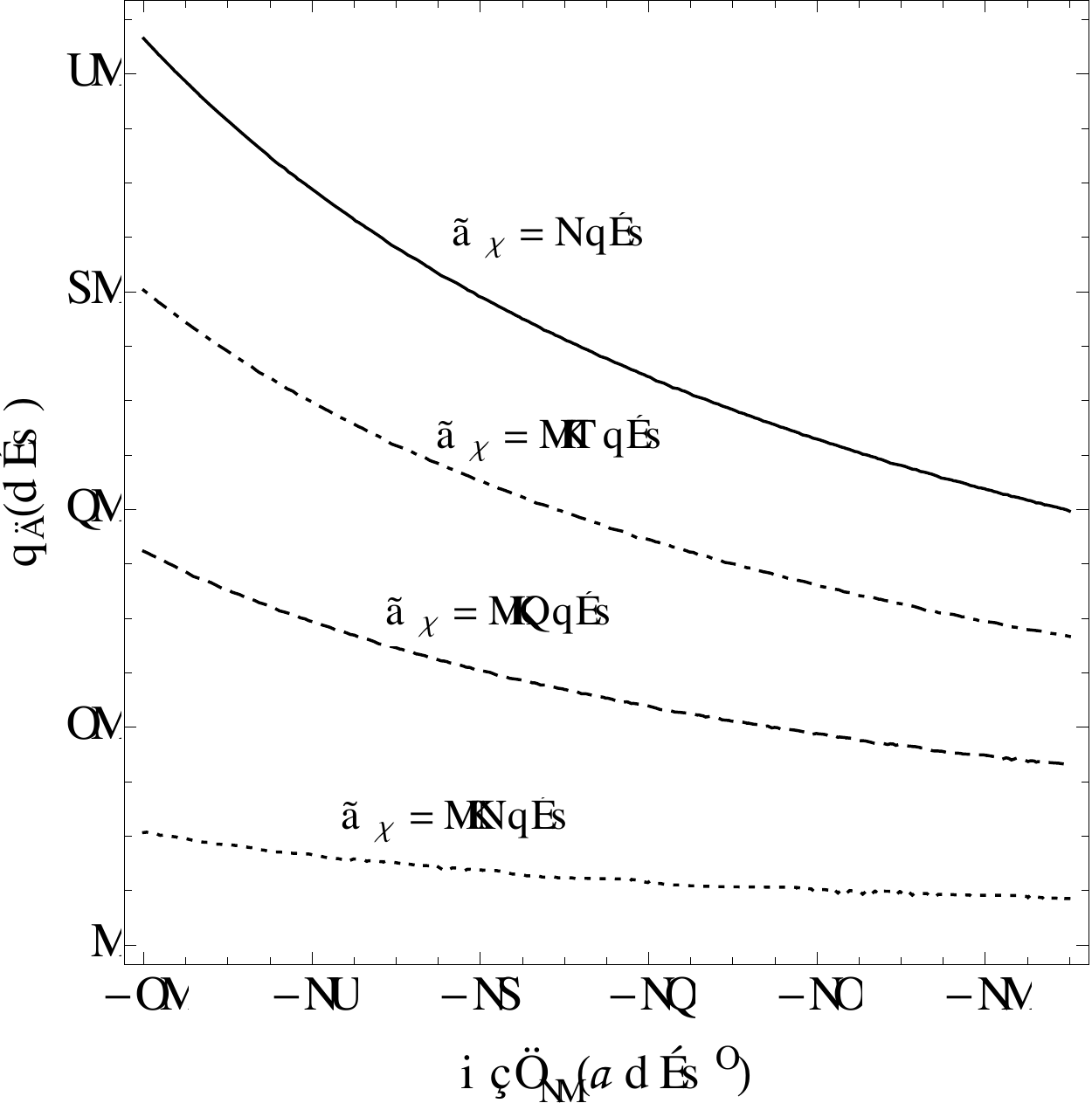}
\caption{$T_b$ as a function of the cross section  $\langle\sigma v \rangle \simeq a$ to reproduce the present relic abundance $\Omega_\chi h^2\simeq0.1$ for various $m_\chi$ in the low-temperature case. 
The ranges of mass and $a$ have been chosen so as to cover a span of popular candidates for cold dark matter.
}\label{fig:fig_low_a-T-m}
\end{figure}

With the help of \eqref{Omega1}  one can explore the relationships between 
$\Omega_\chi h^2$, $x_b$ (or $T_b$), cross section (i.e. $a$ for low-temperature case, while $\sigma_0$ for high-temperature one) and $m_\chi$. Some results are illustrated in  Fig. \ref{fig:fig_low_a-x-Omega} - \ref{fig:fig_low_a-T-m}, where we have chosen $g_*=90$, $g_\chi=g_{\rm eff}=1$ and $n=2$. Note that there is an additional constraint \eqref{NEQ} coming from the requirement of non-chemical equilibrium and the axis ranges in the plots have been chosen so as to satisfy \eqref{NEQ}.  

Fig. \ref{fig:fig_low_a-x-Omega} and \ref{fig:fig_high_a-x-Omega} show contour plots of various relic abundances calculated by \eqref{Omega1} with $m_\chi=100{\rm GeV}$ in low- and high-temperature case respectively.  From these plots one can see that decreasing $x_b$, i.e. raising $T_b$, increases the relic abundance for fixed cross section. On the other hand, for fixed $x_b$ larger cross section predicts larger relic abundance, which is the feature of out-of-equilibrium production, i.e. $\Omega_\chi \propto \langle\sigma v \rangle$.      
Fig. \ref{fig:fig_low_a-x-m} and \ref{fig:fig_high_a-x-m} shows $x_b$ as a function of cross section for various masses in order to reproduce the observed value of the present relic abundance, $\Omega_\chi h^2\simeq0.1$. 
As can be seen from the plots, larger cross section requires larger $x_b$, i.e. lower $T_b$.      

From these considerations one can constrain the temperature $T_b$ for given mass 
and annihilation cross section of dark matter in order to satisfy the observational constraint.  
For instance, 
we provide an estimate of $T_b$ in Fig. \ref{fig:fig_low_a-T-m}, where 
 ranges of  mass (from $0.1$ to $1\, {\rm TeV}$) and annihilation cross section 
(less than weak cross section $\sigma_{\rm weak}\sim 10^{-9}\, {\rm GeV}^{-2}$) have been chosen so as to cover a span of popular candidates for cold dark matter such as supersymmetric neutralino and lightest Kaluza-Klein particle widely studied in the literature
(see e.g. \cite{Bertone:2004pz} for a review). 
In this figure $T_b$ increases with mass and ranges from a few GeV to about 100 GeV in the given range of parameters.



\section{Conclusion}

In this work we have studied a  scenario of dark matter production in the context of bouncing cosmology, in which dark matter was thermally produced from background plasma in the contracting as well as expanding phase. We invoked mechanism of dark matter production in completely out-of-chemical equilibrium since otherwise it would be indistinguishable from standard thermal decoupling scenario. An assumption on bouncing phase has been employed that the bouncing transition was too short to change the number of dark matter particles per comoving volume.  It has been shown that the predicted abundances depend on the temperature at the end/beginning of radiation dominated era in contracting/expanding phase, which can be thought of as a critical temperature of the phase transition from standard physics regime into new physics one, which is responsible for the bounce. Focusing on two illustrative cases, i.e. very low-and high-temperature in comparison with mass of dark matter particle, we have explicitly shown that this temperature ($T_b$) can be constrained by requiring to reproduce the observed present relic abundance of dark matter.   
Our study demonstrates how this alternative route of dark matter production can provide a means of probing a bouncing universe. 
Furthermore, some dark matter models could be discarded if they yield too low $T_b$, e.g. lower than the energy scale of big bang nucleosynthesis ($\sim$ a few MeV), assuming that dark matter was produced out of equilibrium.      


\paragraph{\bf Acknowledgments.}
We would like to thank Robert Brandenberger and Konstantin Savvidy for many useful discussions.
This research project has been supported in parts by the Jiangsu Ministry of Science and Technology under contract BK20131264. 
We also acknowledge 985 Grants from the Ministry of Education, and the Priority Academic Program Development for Jiangsu Higher Education Institutions (PAPD).

\clearpage
\addcontentsline{toc}{section}{References}

\bibliographystyle{JHEP}

\bibliography{DMBUref}

\end{document}